\documentclass[]{spie}  

\usepackage{amsmath,amsfonts,amssymb}
\usepackage{graphicx}
\usepackage[colorlinks=true, allcolors=blue]{hyperref}

\title{No need to modulate: on-sky results of a neural network enhanced pyramid wavefront sensor and prospects for the ELTs}
\author[a,b]{Rico Landman}
\author[b]{Liam Koning}
\author[b,c]{Sebastiaan Y. Haffert}
\author[d]{Joseph D. Long}
\author[c]{Jared R. Males}
\author[b]{Matthijs Mars}
\author[c]{Laird M. Close}
\author[c,e,f, g]{Olivier Guyon}
\author[c]{Warren B. Foster}
\author[c]{Kyle Van Gorkom}
\author[h]{Alexander D. Hedglen}
\author[c]{Parker T. Johnson}
\author[g]{Maggie Y. Kautz}
\author[g]{Jay K. Kueny}
\author[c]{Jialin Li}
\author[g]{Joshua Liberman}
\author[c,e]{Miles Lucas}
\author[g]{Jennifer Lumbres}
\author[g]{Eden A. McEwen}
\author[i]{Avalon McLeod}
\author[j]{Lauren Schatz}
\author[b]{Elena Tonucci}
\author[g]{Katie Twitchell}

\affil[a]{NOVA, Netherlands Research School for Astronomy, 2300 RA Leiden, The Netherlands}
\affil[b]{Leiden Observatory, Leiden University, PO Box 9513, 2300 RA Leiden, The Netherlands}
\affil[c]{Steward Observatory, University of Arizona, 933 North Cherry Avenue, Tucson, AZ, USA}
\affil[d]{Center for Computational Astrophysics, Flatiron Institute, 162 5th Avenue, New York, NY, USA}
\affil[e]{Subaru Telescope, National Astronomical Observatory of Japan, 650 N. A'ohoku Place, Hilo, HI, USA}
\affil[f]{Astrobiology Center, National Institutes of Natural Sciences, 2-21-1 Osawa, Mitaka, Tokyo, Japan}
\affil[g]{Wyant College of Optical Sciences, The University of Arizona, 1630 E University Blvd, Tucson, Arizona, USA}
\affil[h]{Northrop Grumman Corporation, 600 South Hicks Road, Rolling Meadows, Illinois, USA}
\affil[i]{Draper Laboratory, 555 Technology Square, Cambridge, Massachusetts, USA}
\affil[j]{Starfire Optical Range, Kirtland Air Force Base, Albuquerque, New Mexico, USA}

\authorinfo{Further author information: (Send correspondence to R. Landman)\\R. Landman: E-mail: rlandman@strw.leidenuniv.nl}

\begin{document}
\maketitle

\begin{abstract}
One of the main limitations of ground-based extreme adaptive optics systems (XAO) is the balance between the temporal and photon noise error. The unmodulated Pyramid Wavefront Sensor (uPWFS) promises significant gains in sensitivity over its modulated counterpart, but its practical use is limited by its linearity range. Nonlinear reconstructors provide a pathway to recover this dynamic range while preserving the sensitivity of the uPWFS, thereby reducing photon noise and improving contrast. We present the real-time implementation of a Convolutional Neural Network (CNN) reconstructor and show on-sky results with MagAO-X, demonstrating robust and stable correction across diverse atmospheric conditions. Significant gains over default operation are seen in the low and moderate Strehl regimes, while the performance is slightly degraded in the high Strehl regime. We diagnose this in simulation and mainly attribute this to a non-optimized training dataset for the high-Strehl regime, rather than a fundamental limitation of the approach. Furthermore, initial simulations of the NN-enhanced uPWFS for a downscaled version of the Extremely Large Telescope (ELT) show substantial gains for fast petal-piston control. These results demonstrate that NN-enhanced wavefront sensing is a viable technology for future high-contrast instruments.
\end{abstract}


\section{INTRODUCTION}
\label{sec:intro}  
The direct imaging of exoplanets and circumstellar disks is one of the main science drivers for the next generation of Extremely Large Telescopes (ELTs). A primary driver is the detection and characterization of temperate exoplanets in the habitable zones around nearby M-dwarfs, which will be accessible in reflected light with instruments such as the Planetary Camera and Spectrograph (PCS) for the ELT\cite{Kasper2021_PCS} and the Giant Magellan Adaptive Optics eXtreme (GMagAO-X) instrument\cite{Males2024_gmagaox, close2024_gmagaox2024, haffert2024_gmagaox}. These planets are found at small angular separations from their host star, necessitating the use of extreme adaptive optics (XAO) systems and coronagraphs.

\begin{figure}[ht]
    \centering
    \includegraphics[width=0.85\linewidth]{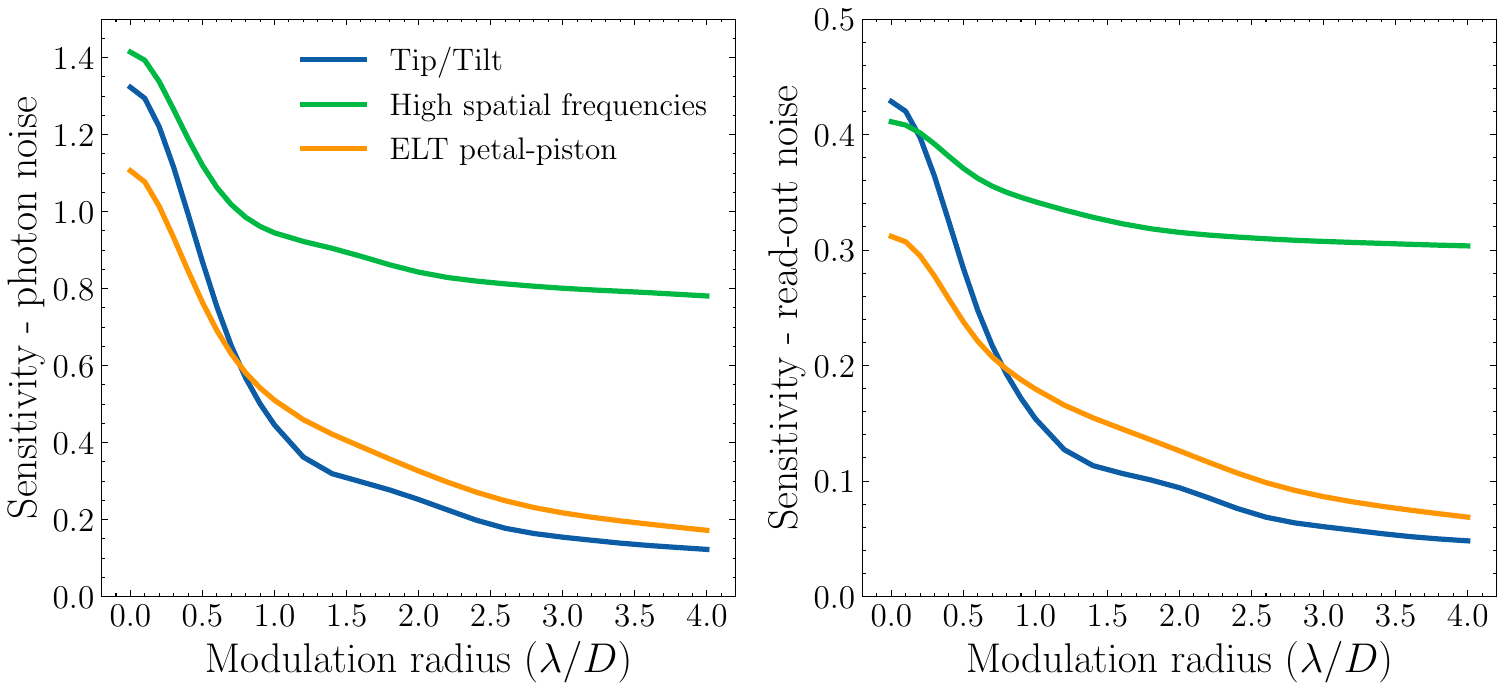}
    \caption{Sensitivity of the PWFS to tip/tilt, high-spatial-frequency, and ELT petal-piston modes as a function of the modulation radius, for the photon-noise-limited (left) and read-noise-limited (right) regimes. The sensitivity to all considered modes is highest for the unmodulated sensor and decreases rapidly with modulation radius, most strongly for the low-order and petal-piston modes.}
    \label{fig:sensitivity}
\end{figure}

At these small separations, the residual wavefront error of an XAO system is dominated by the interplay between the temporal error and the measurement (photon noise) error. Correcting the atmosphere faster reduces the temporal error but leaves fewer photons per wavefront-sensor measurement, increasing the noise error, and vice versa. The pyramid wavefront sensor (PWFS\cite{Ragazzoni1996_pwfs}) is the sensor of choice for most current and planned XAO systems because of its high sensitivity \cite{Ragazzoni1999_pwfs_sensitivity, Correia2020MNRAS_PWFS}. Almost all of these systems, however, operate the PWFS in its modulated form to increase its linearity range, at the cost of a substantial reduction in sensitivity, especially to low-order modes\cite{Chambouleyron2023A&A_noise_propagation_ffwfs, Agapito2023_nonmodulated_pwfs}. Removing the modulation therefore directly improves the sensitivity of the sensor: fewer photons are needed to sense the wavefront, allowing the control loop to run at higher speeds or gains and extending XAO operation to fainter guide stars. Figure~\ref{fig:sensitivity} illustrates this gain, showing the photon-noise and read-noise sensitivity of the PWFS to a range of aberrations as a function of modulation radius.

Removing the modulation has a second important benefit for the segmented apertures of the ELTs. The large secondary spiders and segment gaps of these telescopes introduce discontinuities in the wavefront that give rise to petal-piston modes and low-wind-effect aberrations. The modulated PWFS has a very low sensitivity to these modes and can converge to an incorrect differential-piston solution\cite{Bertrou-Cantou2022_petalpiston_pwfs, Hedglen2022_segment_phasing_pwfs, Engler2022_flipflop}. As a phase sensor, the unmodulated PWFS (uPWFS) is intrinsically much more sensitive to these discontinuities and is a promising way to sense and control them\cite{Levraud2022_petal_elt, Levraud2024_petal_strategy}.

The main obstacle to using the uPWFS is its nonlinear response to large phase aberrations\cite{Esposito2001_pwfs_partial_correction}. Without modulation, the linearity range of the PWFS is much smaller than the aberrations typically imposed by atmospheric turbulence, which makes it difficult to close the loop on-sky with a conventional linear reconstructor. Many approaches have been proposed to overcome the nonlinearity of both the uPWFS and the modulated PWFS, ranging from first-order optical-gain compensation\cite{Deo2019_optical_gain_tracking, Chambouleyron2020_opticalgain, Chambouleyron2021_opticalgain_focalplane} to model-based nonlinear reconstructors\cite{Hutterer2018_hutterer_landweber, Frazin2018_gradient_based_estimation, Hutterer2023_nonlinear_reconstruction_framework, Chambouleyron2023_gs_reconstruction, Haffert2024_zernike_nonlinear} and data-driven reconstructors based on neural networks (NNs)\cite{2018SPIE10703E..1FS_swanson_cnn_pred, Landman2020_nonlinear_cnn, Archinuk2023_nonlinear_reconstruction_ml, Wong2023_nn_reconstruction, Pou2024OExprRL_ML_unmodulated, Weinberger2024A&A}. NNs are an attractive solution because they can approximate the inverse of the nonlinear PWFS response directly from data while preserving the full sensitivity of the sensor. Additionally, they are a software-only solution and do not require the introduction of additional optics.

In Landman et al.\cite{Landman2024} (hereafter Paper~I) we demonstrated closed-loop internal-source laboratory reconstruction of the uPWFS with a convolutional neural network (CNN) on the Magellan Adaptive Optics eXtreme (MagAO-X) instrument, and in Landman et al.\cite{Landman2025} (hereafter Paper~II) we presented the first on-sky demonstration of XAO with an unmodulated PWFS and a nonlinear reconstructor. In these proceedings we briefly recap the method and the on-sky results of Paper~II (Sections~\ref{sec:methods} and \ref{sec:onsky}) and then present new results. On-sky, we extend the demonstration to poorer seeing conditions, showing that the unmodulated PWFS can still operate in this regime. We then present new end-to-end simulations of the MagAO-X performance as a function of guide-star magnitude, and study the possible limitations of the on-sky performance. Finally, we show in simulations that the Neural Network-enhanced uPWFS can sense and control petal-piston modes on a downscaled model of an ELT aperture (Section~\ref{sec:simulations}). We close with an outlook towards deployment on next-generation high-contrast instruments (Section~\ref{sec:outlook}).

\section{METHODS}
\label{sec:methods}
Our reconstructor and its real-time implementation are described in detail in Paper~I and Paper~II; we summarize them here. The CNN follows a U-Net architecture\cite{2015arXiv150504597R_unet} with a nonlinear encoder--decoder with skip connections and a direct linear connection between the input and output that allows the model to reproduce a linear reconstructor \cite{Landman2020_nonlinear_cnn}. The input consists of the four reference-subtracted and normalized PWFS pupil images, each sampled at $64\times64$ pixels and concatenated along the channel dimension. The network outputs the coefficients of the controlled modes, which are then projected onto the deformable mirror (DM) actuators. On MagAO-X we reconstruct and control the standard set of 1563 tweeter modes.

The training data are generated using the internal source. Random phase screens with random power-law power spectral densities (power-law index between -1 and -3) are applied to the DM and the corresponding PWFS images are recorded, spanning a wide range of RMS wavefront error (from $\sim0.2$ to $600$~nm) and guide-star magnitudes, so that the model performs well over the full range of aberrations encountered when closing and maintaining the loop. The network is trained to minimize a relative RMS loss, which weights small and large aberrations equally and is essential for good closed-loop behaviour\cite{Landman2020_nonlinear_cnn}. We refer to Paper II for more details about the calibration procedure. 

\begin{figure}[ht]
    \centering
    \includegraphics[width=\linewidth]{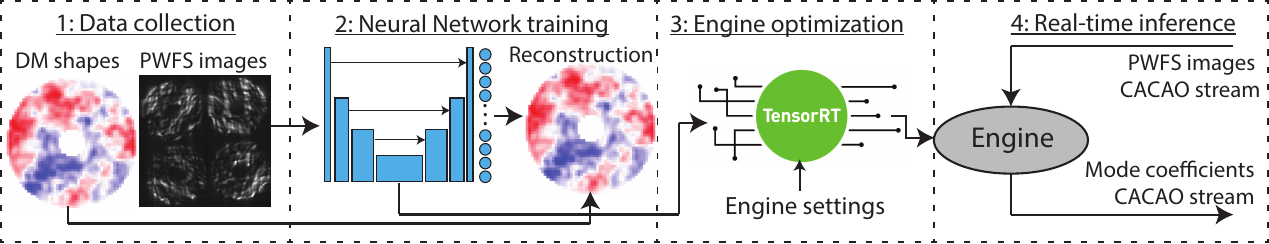}
    \caption{Overview of the pipeline and real-time implementation of the CNN reconstructor, adapted from Landman et al.\cite{Landman2025} Training data (DM shapes and PWFS images) are collected on the internal source, the CNN is trained on these data, converted to an optimized TensorRT engine, and finally run in real time within the MagAO-X control software, reading the PWFS image stream and writing the reconstructed mode coefficients.}
    \label{fig:pipeline}
\end{figure}

To run the control loop at the multiple-kilohertz frequencies required by MagAO-X, the trained model is converted to an optimized inference engine with NVIDIA TensorRT and implemented in the eXtreme Wavefront Control Toolkit (XWCTk) \cite{males_spie_2026_xwct}, which uses the CACAO real-time control framework\cite{Guyon2018_cacao, deo_spie_2026_cacao}. The application reads the reference-subtracted and normalized PWFS image stream, propagates it through the optimized TensorRT engine, and writes the resulting mode coefficients back to the stream used by the real-time controller. The full pipeline is summarized in Figure~\ref{fig:pipeline}. On the RTX~4090 GPU in MagAO-X the inference latency is below $250~\mu$s at single precision and below $125~\mu$s at half precision. We now run the engine at half precision, which has a negligible impact on the reconstruction accuracy while comfortably supporting loop speeds above 2~kHz. The intermittent latency jitter reported in Paper~II, which was traced to an issue with the DM kernel module at high loop speeds, has since been resolved.

\section{ON-SKY RESULTS}
\label{sec:onsky}
The first on-sky tests of the smart uPWFS were carried out with MagAO-X on the 6.5~m Magellan Clay telescope and were presented in Paper~II; here we extend that demonstration to poorer seeing conditions and to a fainter guide star. To enable a fair comparison, we first closed the loop and tuned the modal gains with the conventional modulated PWFS and linear reconstructor, and then switched to the unmodulated PWFS with the CNN reconstructor while keeping the same operational settings. This switching was repeated to verify that the atmospheric conditions did not change significantly between tests. The Strehl ratio was estimated from the focal-plane images of the science camera.

\begin{figure}[ht]
    \centering
    \includegraphics[width=\linewidth]{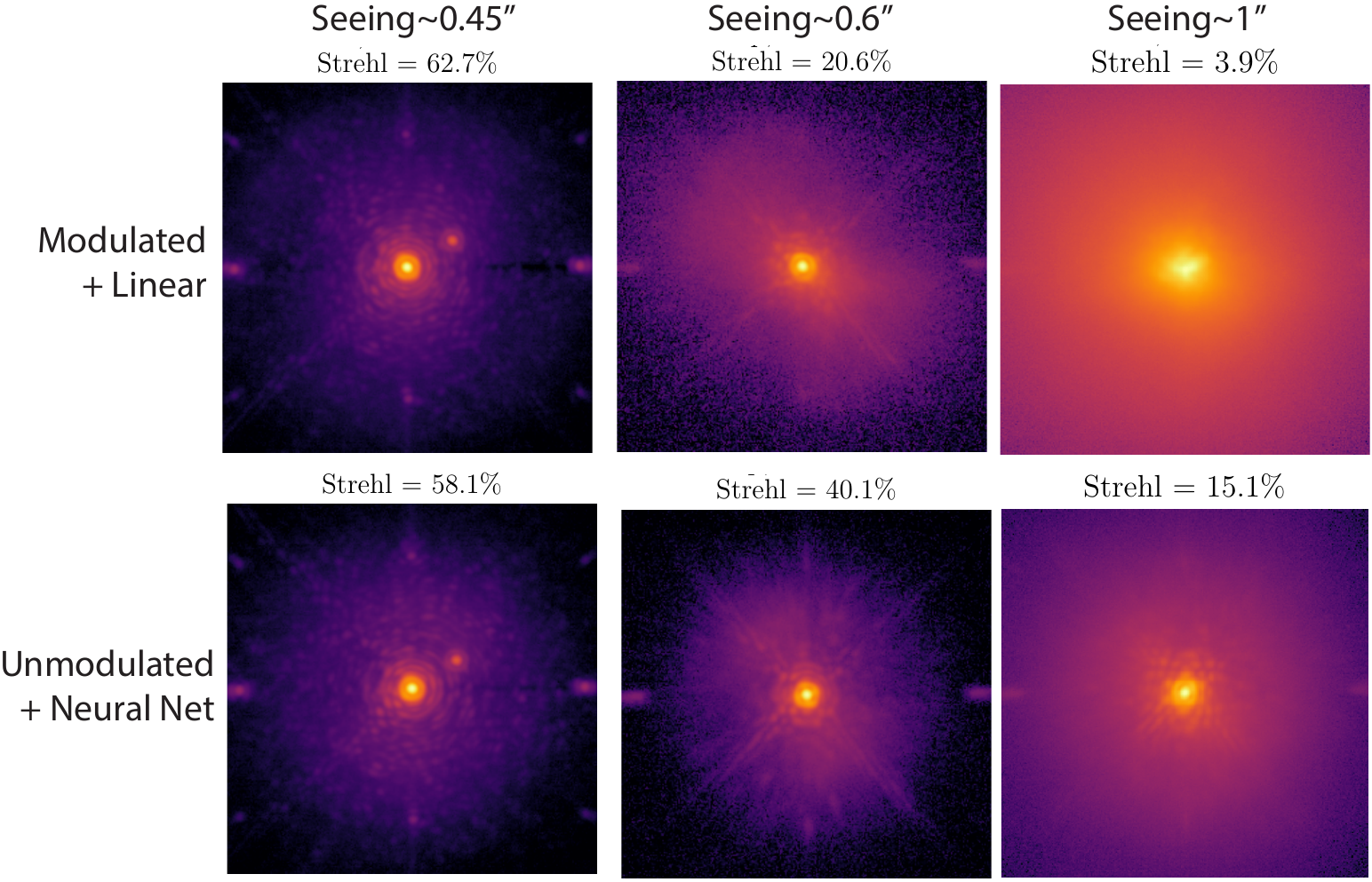}
    \caption{On-sky integrated PSFs comparing the default modulated PWFS with linear reconstruction (top) to the unmodulated PWFS with the CNN reconstructor (bottom), for three targets under increasing seeing (left to right).}
    \label{fig:onsky}
\end{figure}

Figure~\ref{fig:onsky} shows the integrated on-sky point spread functions (PSFs) for three representative targets spanning a range of seeing conditions, comparing the default modulated PWFS with linear reconstruction (top row) to the unmodulated PWFS with the CNN reconstructor (bottom row). In the best seeing ($\sim0.45''$, target $\alpha$~Eri), the CNN reaches a Strehl ratio of 58.1\%, slightly below the 62.7\% obtained with the modulated PWFS. As the seeing degrades, however, the unmodulated sensor with the CNN gives increasingly better performance over default operation. At $\sim0.6''$ seeing with relatively strong winds (AF~Lep), the CNN reaches 40.1\% compared to 20.6\% for the modulated PWFS, while in poor seeing of $\sim1''$ (HD~15115) it achieves 15.1\% versus only 3.9\%. This is the result of the instantaneous wavefront frequently exceeding the linearity range of the sensor under poor seeing conditions, even for the modulated PWFS. As a result, the linear reconstructor underestimates these large aberrations and struggles to converge, whereas the nonlinear CNN continues to reconstruct them more accurately, maintaining a higher level of correction.

\subsection{Performance on a fainter star}
\begin{figure}[ht]
    \centering
    \includegraphics[width=0.8\linewidth]{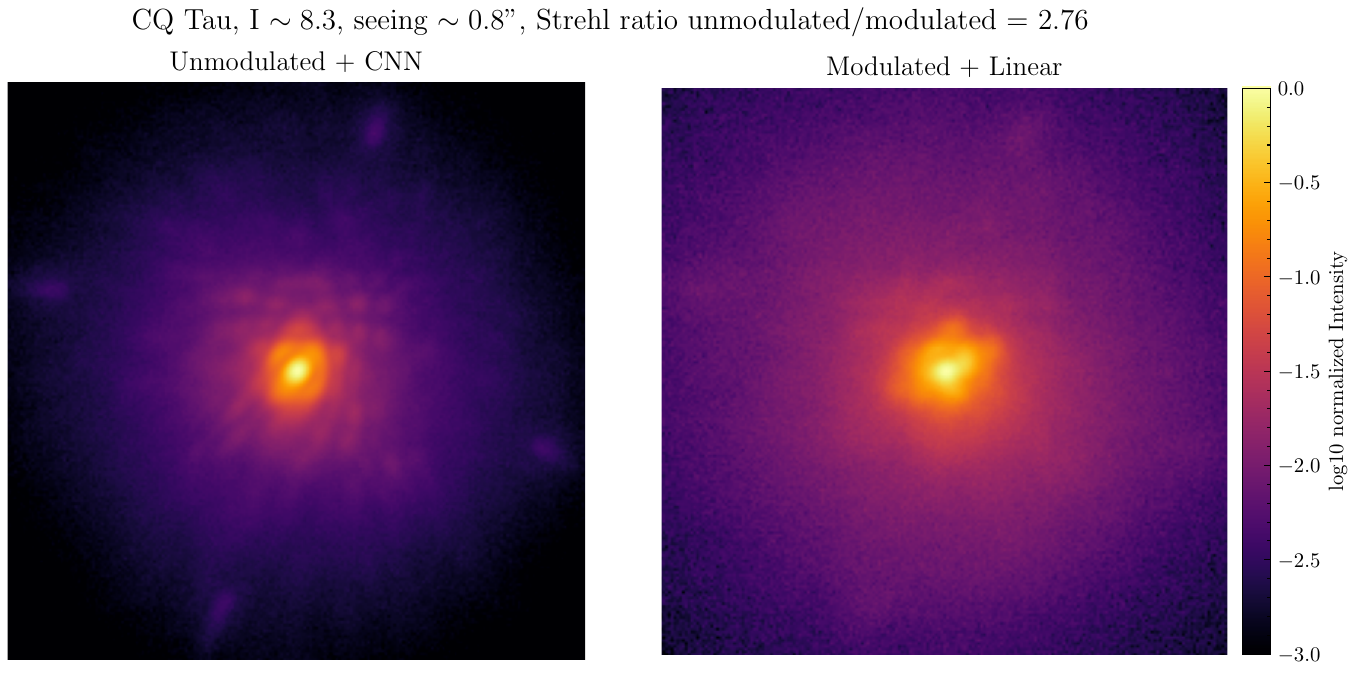}
    \caption{On-sky PSFs comparing the modulated PWFS with linear reconstruction (right) to the unmodulated PWFS with the CNN reconstructor (left) on a fainter star ($I\sim8.3$), showing the significantly improved performance from the unmodulated PWFS.}
    \label{fig:faint}
\end{figure}

The main motivation for removing the modulation is the improved sensitivity, which is expected to yield the largest gains on faint guide stars, where the measurement is photon limited. Figure~\ref{fig:faint} compares the modulated PWFS with linear reconstruction to the unmodulated PWFS with the CNN reconstructor on a slightly fainter star ($I\sim8.3$). While this is still not a very faint target, photon noise will start to contribute to the error budget at 2 kHz. We observe a significant improvement in image quality with the unmodulated sensor. However, the seeing during these observations was also suboptimal ($\sim 0.8"$), so the improvement likely stems predominantly from the better handling of large aberrations by the nonlinear reconstructor, as opposed to the improved sensitivity. Disentangling the two contributions will require observations of an $I\sim10$ star under good seeing, which we hope to do on a future run.

\subsection{Bright star in good seeing}
\begin{figure}[ht]
    \centering
    \includegraphics[width=0.8\linewidth]{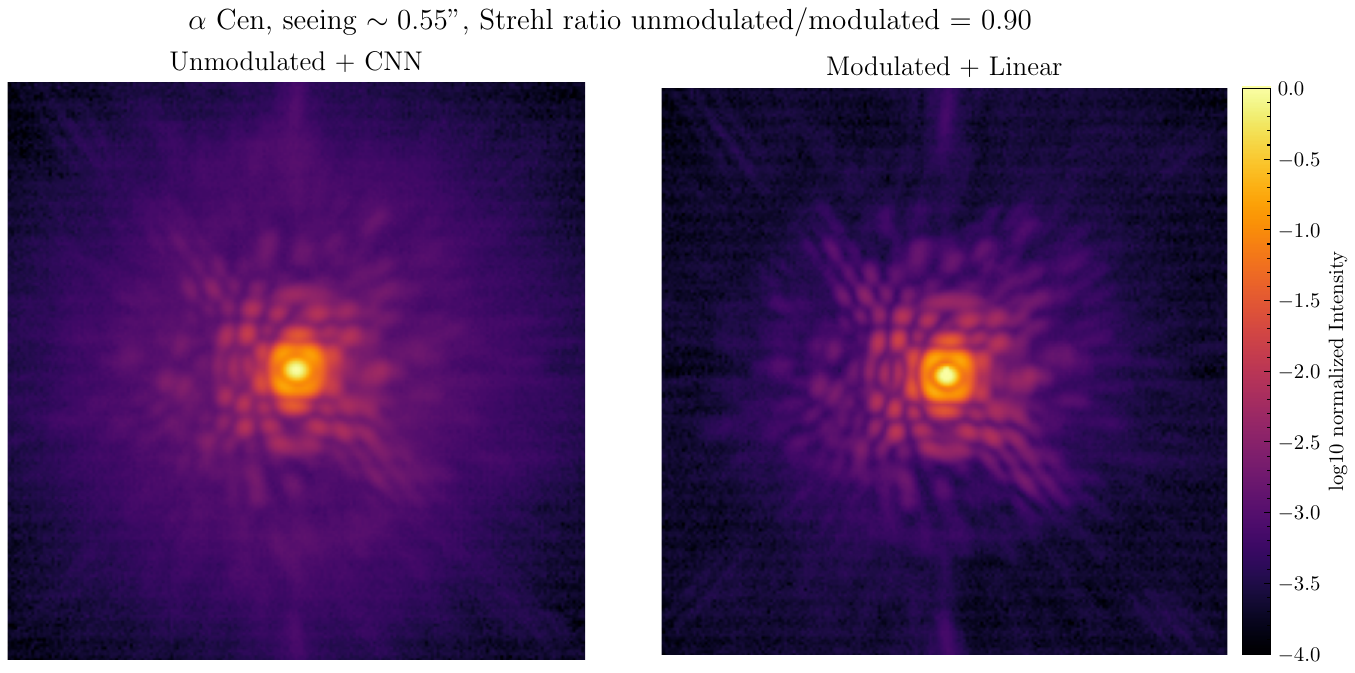}
    \caption{On-sky PSFs comparing the modulated PWFS with linear reconstruction (right) to the unmodulated PWFS with the CNN reconstructor (left) on a bright star under good seeing conditions, showing the decreased obtained performance from the unmodulated PWFS.}
    \label{fig:alphcen}
\end{figure}

We again tested the performance of our approach on a bright star in good seeing conditions. Compared to Paper II, we had hoped that the improved calibration approach, which was sped up by a factor 5, as well as the DM kernel fix, would improve the performance in this regime. Fig. \ref{fig:alphcen} shows the results during the most recent observing run in Spring 2026. The CNN controller, however, still performs worse than the default modulated PWFS operation. Although the difference in Strehl ratio is modest, the degradation in contrast is substantial, as illustrated by the long-exposure images in Fig.~\ref{fig:alphcen}. Several effects may limit the performance in this high-Strehl regime. These effects are investigated through simulations in Section~\ref{sec:sim_diag}.

\section{SIMULATIONS: MagAO-X}
\label{sec:simulations}
To understand the on-sky behaviour and to explore the potential of the smart uPWFS beyond the current MagAO-X tests, we performed a series of end-to-end simulations, both for MagAO-X and a downscaled version of the ELT aperture. The atmosphere is simulated with \texttt{hcipy}\cite{por2018hcipy} using turbulence statistics representative of Las Campanas Observatory\cite{prieto2010giant, thomas2010giant, males2018ground}, and the closed loop is run at 2~kHz with an effective 1.5-frame delay. The sensing wavelength is $875$~nm, the $6.5$~m pupil is sampled on a $128\times128$ grid, and the same 1563 controlled modes and 2500 tweeter actuators are used as on-sky. We used a scalar gain for all modes that was re-optimized for each stellar magnitude and for each reconstructor, and all comparisons within a figure use the same atmospheric realizations for every reconstructor.

\subsection{MagAO-X: Strehl vs. magnitude}
\label{sec:sim_magaox}
\begin{figure}[htbp]
    \centering
    \includegraphics[width=0.7\linewidth]{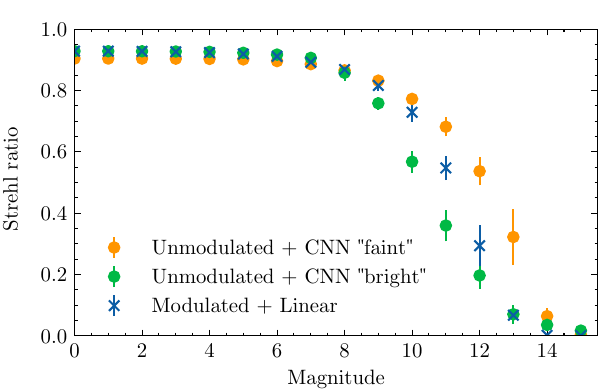}
    \caption{Simulated closed-loop Strehl ratio as a function of guide-star magnitude for the modulated PWFS with linear reconstruction and the unmodulated PWFS with the CNN reconstructor.}
    \label{fig:strehl_mag}
\end{figure}

Figure~\ref{fig:strehl_mag} shows the simulated closed-loop Strehl ratio as a function of guide-star magnitude for the modulated PWFS with linear reconstruction and the unmodulated PWFS with the CNN reconstructor. Here, we show the results of two different reconstructors: A "bright/high-Strehl" model which was only trained on magnitudes $I<8$, with a dataset containing smaller RMS phase screens and flatter power-laws, and a "faint/low-Strehl" model, which was trained on data with magnitudes $I<10$ with more high RMS phase screens and steeper power-laws. The faint-star model is trained in the same way as the on-sky models were generally trained. For bright stars, where the correction is mainly limited by the temporal and fitting errors rather than by noise, the three reconstructors achieve similar performance in terms of Strehl, with a decrease in Strehl of few percent for the faint-star model. This is in good agreement with our on-sky results, where we also see a slight decrease in performance in the high-Strehl regime. While this slight decrease in Strehl seems minimal, it can have detrimental effects on the contrast. To study what is causing this decreased performance, we have run a number of tests, which will be discussed in the next subsection.

Towards fainter magnitudes the faint-star model demonstrates the higher sensitivity of the unmodulated sensor: beyond $I\sim8$ it maintains a systematically higher Strehl ratio and extends the limiting magnitude at which useful correction can be obtained by about a full magnitude.

\subsection{What limits the bright-star performance?}
\label{sec:sim_diag}
The closed-loop Strehl ratio provides an overall measure of AO performance, but does not distinguish between reconstruction bias, photon-noise propagation, and control-loop effects. Especially for a nonlinear reconstructor, these effects are harder to distinguish from each other. To isolate the reconstruction itself, we performed a series of open-loop experiments. A known closed-loop residual wavefront error is propagated through the WFS and the wavefront is reconstructed. Rather than using the power-law phase screens employed during training, we evaluate the reconstructors on snapshots of the residual wavefronts obtained during closed-loop operation, as these are more representative of the wavefronts encountered during AO operation. 

\subsubsection{Reconstruction bias}
\begin{figure}[htbp]
\centering
\includegraphics[width=0.55\linewidth]{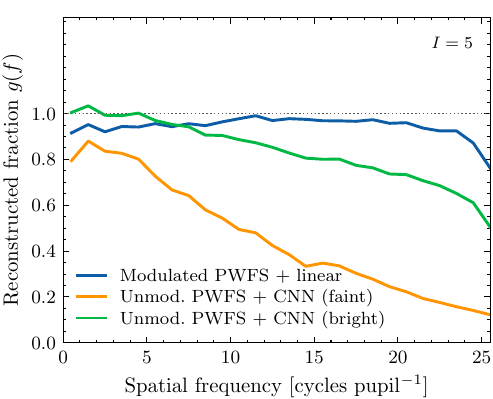}
\caption{Reconstructed fraction of true closed-loop residuals, showing the decreased implicit modal gain that occurs for the CNNs due its bias from training on power-law errors and noisy data.}
\label{fig:gain_f}
\end{figure}

Figure~\ref{fig:gain_f} quantifies the implicit gain of the reconstructor as a function of spatial frequency. For each spatial frequency we compute the slope of a least-squares fit between the reconstructed and true Fourier coefficients for a range of closed-loop residuals. We refer to this slope as the reconstructed fraction $g(f)$. Values of $g=1$ correspond to perfect reconstruction, whereas $g<1$ indicates underestimation. The modulated PWFS with the linear reconstructor recovers approximately a flat $0.9-1$ fraction of the wavefront across the full controlled spatial-frequency range. In contrast, for the ``faint'' CNN, $g(f)$ drops significantly for higher spatial frequencies, which means it underestimates higher-order modes. This likely originates from the training data, where the relative RMS loss and the predominance of power-law phase screens bias the network towards underestimating the small, spectrally flat closed-loop AO residuals. Retraining the network on a narrower magnitude range and with smaller RMS phase screens largely removes this bias, with the ``bright'' model reaching a much flatter reconstructed fraction. This reduced effective modal gain could potentially be compensated through self-tuning temporal control algorithms.

\subsubsection{Noise propagation}
\begin{figure}[htbp]
    \centering
    \includegraphics[width=\linewidth]{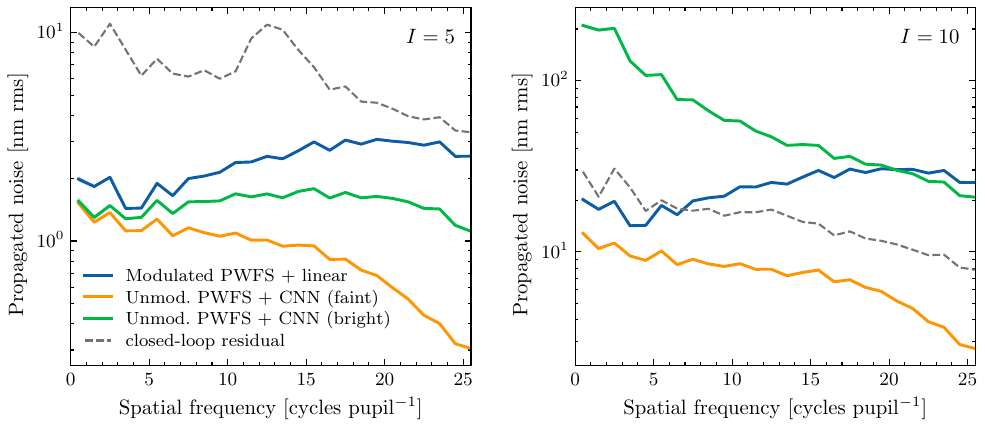}
    \caption{Noise propagation from photon noise to the reconstructed wavefront for a magnitude $I=5$ star (left) and $I=10$ (right). This is tested on closed-loop residuals and considers all effects, including for example optical gains and nonlinear noise amplification.}
    \label{fig:noise_prop}
\end{figure}

Figure~\ref{fig:noise_prop} shows the noise propagation of each reconstructor, evaluated by repeatedly sensing identical closed-loop residuals with independent photon-noise realizations and measuring the standard deviation of the reconstructed wavefront. This procedure naturally includes the nonlinear response of the PWFS, including optical gain variations and potential nonlinear noise amplification around its closed-loop operating point. At $I=5$, both CNNs propagate less noise than the modulated PWFS over the full controlled spatial bandwidth. At $I=10$, the expected sensitivity gain becomes evident: the ``faint'' model has significantly decreased noise propagation compared to the modulated PWFS. The ``bright'' model, however, strongly amplifies the noise and becomes unstable, illustrating the nonlinear noise amplification that can happen if a model is not properly regularized. These results illustrate the familiar bias--variance trade-off. Relative to the modulated PWFS, the ``faint'' CNN introduces additional reconstruction bias while substantially reducing propagated measurement noise. Retraining shifts this balance towards lower bias at the expense of increased noise propagation. The reduced noise propagation does mean that the unmodulated PWFS can likely be run at higher gains and/or faster loop speeds.

\subsubsection{Closed-loop convergence}
\begin{figure}[htbp]
    \centering
    \includegraphics[width=\linewidth]{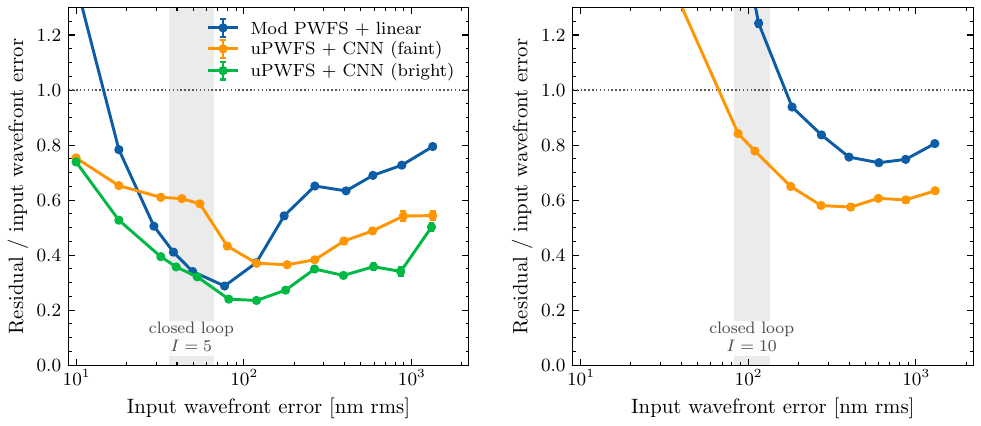}
    \caption{Relative reconstruction error on true simulated closed-loop runs for guide star magnitudes I=5 (left) and I=10 (right). The typical closed-loop RMS regime for the nominal seeing and this magnitude is indicated by the grey shaded region.}
    \label{fig:amplitude}
\end{figure}

Figure~\ref{fig:amplitude} shows how the reconstruction accuracy of the residual wavefront evolves throughout closed-loop convergence under seeing conditions between $0.5''$ and $0.8''$. This produces realistic residuals spanning the full range from the uncorrected atmosphere to the final operating point. As the loop closes, we move from large open-loop turbulence (right side of the plot), to the closed-loop operating point (left side of the plot), while the spatial distribution of the residual wavefront changes. All reconstructors recover a substantial fraction of the wavefront error during the closing of the loop, but their convergence rates differ significantly. At $I=5$, the unmodulated PWFS with the ``bright'' CNN correctly estimates $64$--$76\%$ of the wavefront error per measurement between $80$ and $900$~nm RMS, compared to $27$--$63\%$ for the modulated PWFS. This improved estimation over a large dynamic range illustrates the nonlinear reconstructor's ability to operate under worse seeing conditions. The  `faint'' CNN estimates the initial phase screens quite well, but has significantly decreased performance around the closed-loop operating point. At $I=10$, the ``faint'' CNN estimates the wavefront more accurately than the modulated PWFS across the full amplitude range, showing its improved sensitivity.

These experiments demonstrate that the reduced performance observed on bright stars is not an intrinsic limitation of either the unmodulated PWFS or the nonlinear reconstructor. Instead, it is caused by a bias introduced by the training distribution, which is directly visible in the reconstructed fraction and largely disappears after retraining on data that is more representative of the high-Strehl regime. At the same time, the bright and faint models occupy different points in the bias--variance trade-off and therefore excel in different magnitude regimes. Their performance crosses around $I\sim7$, suggesting that switching between dedicated bright- and faint-star reconstructors is preferable to using a single network across the full magnitude range.

\subsubsection{Dispersion on the tip of the PWFS}
\begin{figure}[htbp]
    \centering
    \includegraphics[width=0.5\linewidth]{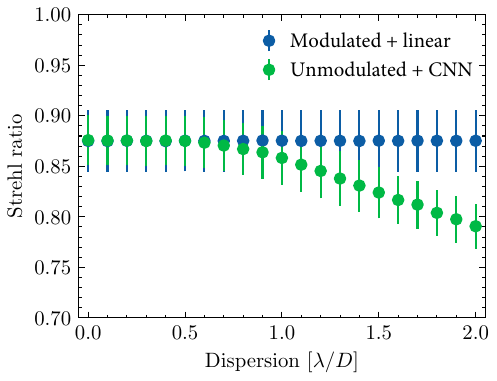}
    \caption{Simulated closed-loop Strehl ratio as a function of the residual dispersion on the tip of the PWFS.}
    \label{fig:dispersion_sim}
\end{figure}

A final, secondary effect that could be limiting our high-Strehl performance on-sky is residual dispersion on the tip of the pyramid, which is currently not cancelled out in MagAO-X for the WFS arm \cite{twitchell_spie_2026_adc}. Figure~\ref{fig:dispersion_sim} quantifies the impact of this dispersion on the obtained Strehl ratio. We simulate the residual atmospheric dispersion as a one-dimensional, diagonal smearing of the PSF at the tip of the pyramid and evaluate the closed-loop Strehl ratio as a function of the amount of dispersion. The modulated PWFS is essentially unaffected, since its $3\lambda/D$ modulation averages over the dispersed beam. The unmodulated PWFS with the CNN, on the other hand, loses a couple of percent of Strehl as the dispersion increases past $\sim1\,\lambda/D$, as the training data did not include this effect.

\section{Simulations: ELT petal-piston control}
\label{sec:sim_elt}
The segmented apertures and/or thick spiders of the ELTs introduce petal-piston modes that are particularly challenging to sense with a modulated PWFS. Because the unmodulated PWFS acts as a phase sensor, it is intrinsically far more sensitive to these discontinuities\cite{Levraud2022_petal_elt, Levraud2024_petal_strategy}, and combining it with a nonlinear CNN reconstructor offers a way to sense and correct petal-piston modes in the presence of atmospheric residuals. To test this, we constructed a downscaled toy model of the ELT aperture: the aperture is scaled to 6.5~m and sampled at $128\times128$ pixels, retaining the dominant spider-induced discontinuities and six independent petal regions. The five independent petal-piston modes are added to the mode basis, and a CNN is trained as before, now with petal-piston modes injected on top of the atmospheric residuals.

\begin{figure}[ht]
    \centering
    \includegraphics[width=\linewidth]{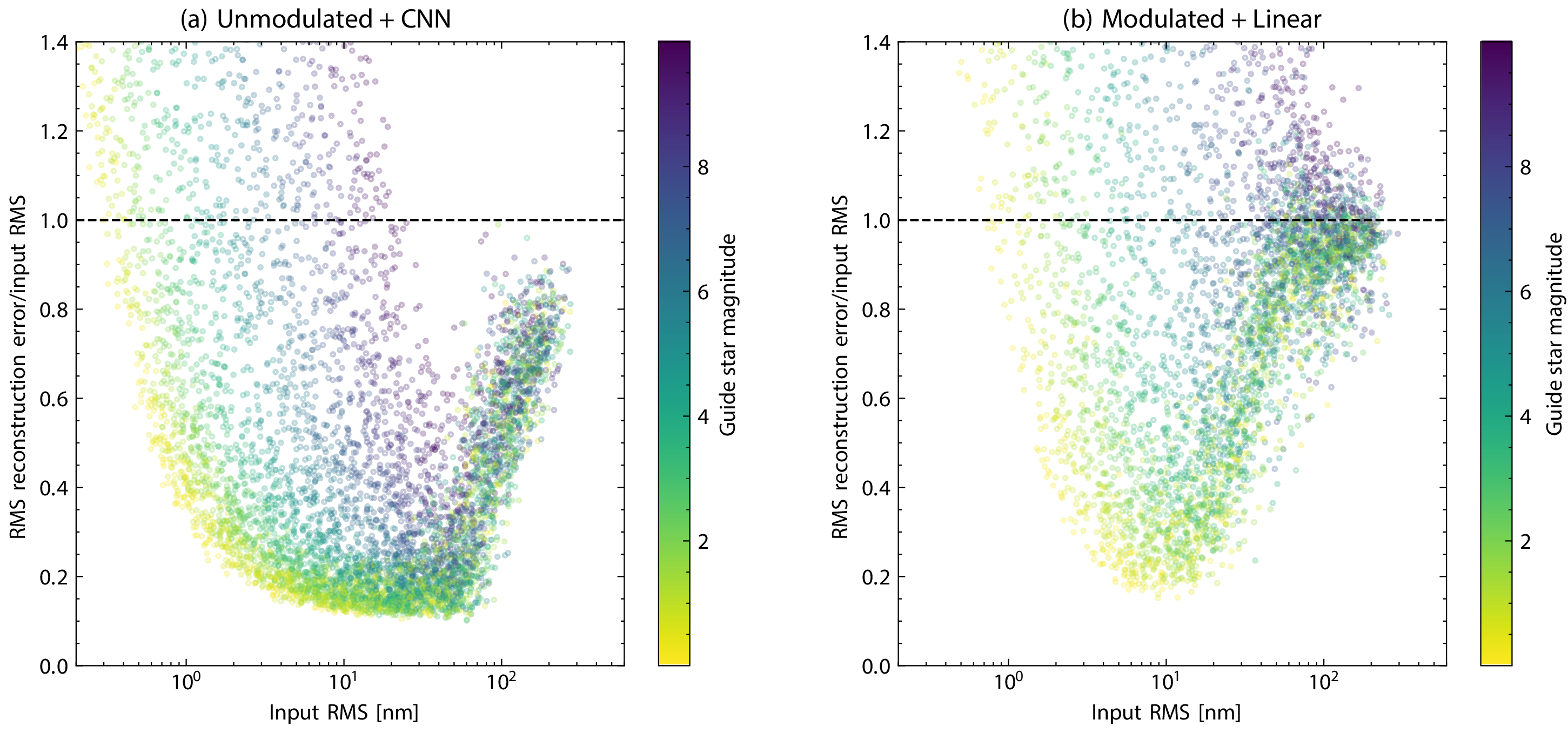}
    \caption{Open-loop reconstruction accuracy on the downscaled ELT aperture in the presence of petal-piston modes, shown as the relative reconstruction error (residual RMS divided by input RMS) as a function of the input RMS wavefront error and colour-coded by guide-star magnitude. The unmodulated PWFS with the CNN (a) achieves a lower relative error, and hence a larger dynamic range, than the modulated PWFS with the linear reconstructor (b), especially for the largest aberrations.}
    \label{fig:elt_recon}
\end{figure}

Figure~\ref{fig:elt_recon} compares the open-loop reconstruction accuracy of the petal-piston modes for the two approaches, showing the relative reconstruction error (residual RMS divided by input RMS) as a function of the input RMS wavefront error, color-coded by guide-star magnitude. The unmodulated PWFS with the CNN (left) reconstructs the petal-piston modes more accurately than the modulated PWFS with the linear reconstructor (right) and over a much larger range. It demonstrates the increased sensitivity and dynamic range, as the reconstruction remains accurate over a wider range of stellar magnitudes and aberration amplitudes.

\begin{figure}[ht]
    \centering
    \includegraphics[width=0.69\linewidth]{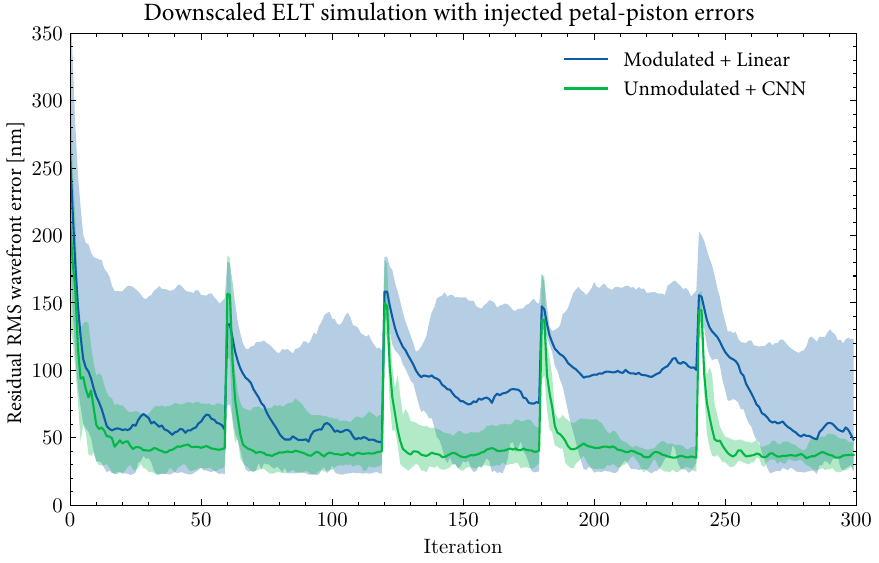}
    \includegraphics[width=0.3\linewidth]{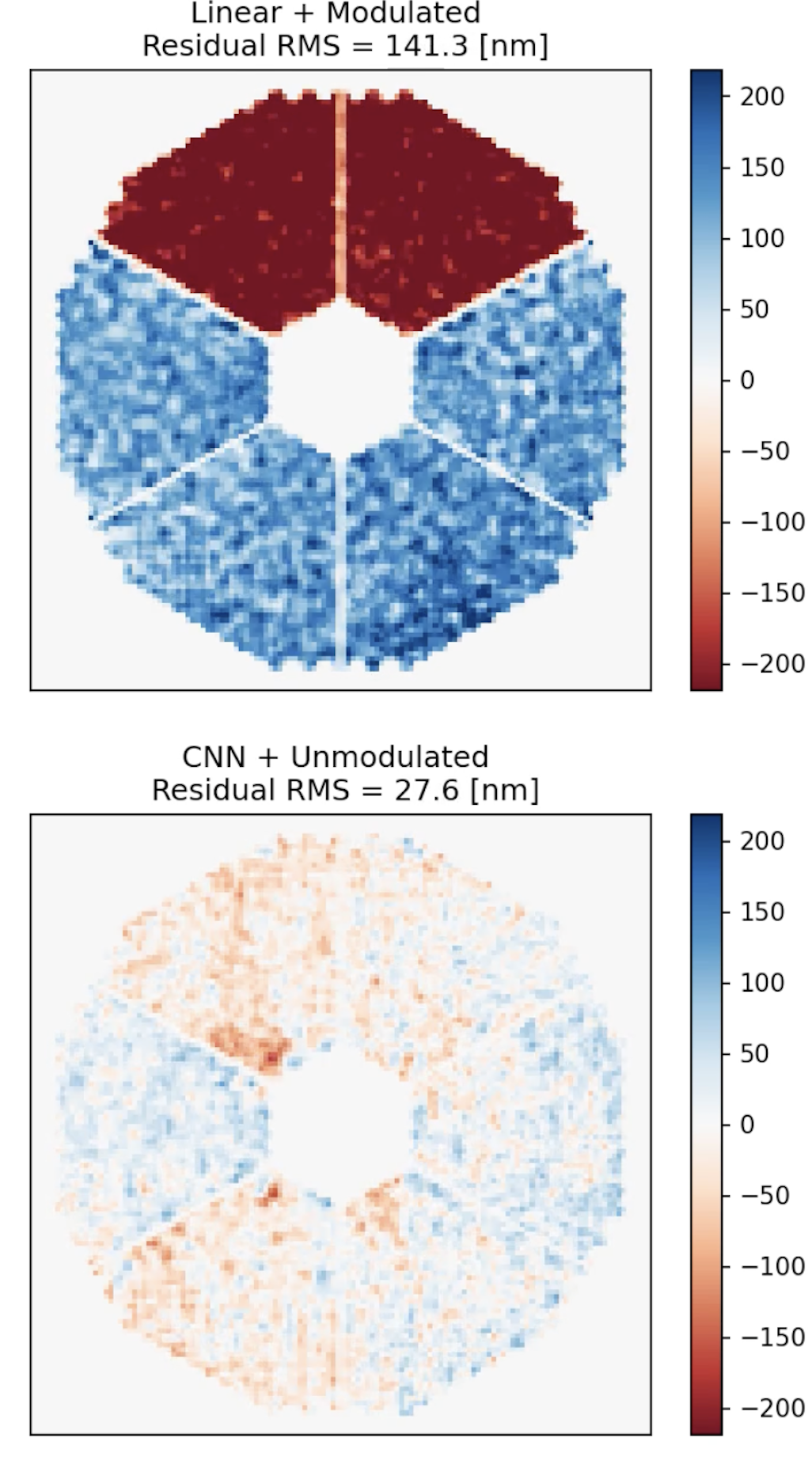}
    \caption{Left: residual RMS wavefront error during a closed-loop simulation on the downscaled ELT aperture and an 8th-magnitude star, where random petal-piston disturbances are injected every 60 iterations. The unmodulated PWFS with the CNN (green) converges faster and to a lower residual than the modulated PWFS with linear reconstruction (blue). Right: a representative residual wavefront error frame, showing that the modulated PWFS with the linear reconstructor leaves a large petal-piston residual with the segments locked at different piston values, while the unmodulated PWFS with the CNN corrects it.}
    \label{fig:elt_closed}
\end{figure}

Figure~\ref{fig:elt_closed} shows a closed-loop simulation in which a random combination of petal-piston modes is injected every 60 iterations (left panel) on top of the atmospheric turbulence. The unmodulated PWFS with the CNN consistently converges faster and to a lower residual wavefront error than the modulated PWFS with the linear reconstructor. The modulated PWFS is not only slower to recover but frequently fails to fully correct the petal-piston aberration. This is illustrated by the representative reconstructed frame in the right panel, where the modulated PWFS with the linear reconstructor leaves a large petal-piston residual, with the segments locked at different piston values. These simulations show that the smart uPWFS can resolve petal-piston modes even in the presence of atmospheric turbulence and demonstrate its potential for petal-piston control on the ELTs.

\section{OUTLOOK}
\label{sec:outlook}
We have demonstrated that an unmodulated PWFS combined with a convolutional neural network reconstructor can provide robust and stable XAO correction on-sky over a wide range of atmospheric conditions. By recovering the dynamic range of the sensor while preserving its intrinsic sensitivity, the smart uPWFS enables operation without modulation and currently shows clear advantages under poorer seeing conditions and in regimes where large residual aberrations dominate the error budget. Simulations have shown that the reduced performance observed in the high-Strehl regime is not an intrinsic limitation of nonlinear reconstruction, but is primarily caused by a non-optimized training distribution of the CNN. Our simulations further demonstrate the potential of this approach for future segmented telescopes, where the increased sensitivity of the uPWFS enables substantially improved petal-piston sensing and control.

Several improvements are already underway. We now operate the CNN reconstructor at half precision and loop speeds above 3~kHz are feasible. Our simulations indicate that training specialized models for different operating regimes can recover the small performance loss observed on the brightest stars, while maintaining the sensitivity advantage at faint magnitudes. We plan to adapt this for the on-sky runs. In addition, the ongoing reduction of internal bench turbulence in MagAO-X will improve calibration stability by reducing sensitivity to slow instrumental drifts. We are also investigating the combination of the nonlinear reconstructor with self-optimizing control laws, such as data-driven subspace predictive control\cite{haffert2021_ddspc} and reinforcement-learning-based predictive control\cite{Nousiainen2022_rl}, to automatically tune the modal gains and further reduce the temporal error.

The favourable computational scaling of CNN inference, which grows approximately as $D^2$ compared to the $D^4$ scaling of dense matrix-vector multiplication\cite{Landman2025}, together with continued improvements in GPU hardware, makes NN-enhanced wavefront sensing a promising approach for ELT-scale systems. We are extending these simulations to full-scale implementations for future high-contrast instruments such as PCS\cite{Kasper2021_PCS, correia_spie_2026_pcs} and GMagAO-X\cite{Males2024_gmagaox, close2024_gmagaox2024, haffert2024_gmagaox}. The combination of improved photon sensitivity and better segment/petal-piston wavefront control could increase the accessible target sample for ELT high-contrast imaging while improving performance on the brightest targets. These results demonstrate that the neural-network-enhanced unmodulated Pyramid wavefront sensor is a viable wavefront sensor for next-generation XAO systems.

\acknowledgments 
We are very grateful for support from the NSF MRI Award \#1625441. The Phase II upgrade program is made possible by the generous support of the Heising-Simons Foundation. MagAO-X uses the CACAO software package, which is supported by NSF Award \#2410616. RL, SYH and MM acknowledge support from NWO Award 184.036.004

\bibliography{report} 
\bibliographystyle{spiebib} 

\end{document}